\documentclass[]{mandm}

\usepackage[utf8]{inputenc}
\usepackage[T1]{fontenc}
\usepackage{graphicx}
\usepackage[hyphens]{url}
\usepackage{hyperref}
\usepackage{orcidlink}
\usepackage{amsmath}
\usepackage{booktabs}
\usepackage{tabularx}
\usepackage{siunitx}

\usepackage[version=4]{mhchem}

\DeclareSIUnit\angstrom{\text {Å}}

\newcommand*\circnum[1]{\raisebox{.5pt}{\textcircled{\raisebox{-.9pt} {#1}}}}

\jourvolume{XX}
\jourissue{Y}
\jourpubyear{2022}

\begin{document}

\sloppy

\title{Phase Object Reconstruction for 4D-STEM using Deep Learning}

\author[T Friedrich, CP Yu et al]{%
  Thomas Friedrich\orcidlink{0000-0002-7584-2080}$^{1,2}$,
  Chu-Ping Yu\orcidlink{0000-0003-1563-5095}$^{1,2}$,
  Johan Verbeeck\orcidlink{0000-0002-7151-8101}$^{1,2}$
  and Sandra Van Aert\orcidlink{0000-0001-9603-8764}$^{1,2}$}

\affiliation{%
  $^1$EMAT, University of Antwerp, Antwerp, Belgium\\
  $^2$NANOlab Center of Excellence, University of Antwerp, Antwerp, Belgium\\
  Corresponding Author: Sandra Van Aert \email{sandra.vanaert@uantwerpen.be}}

\begin{frontmatter}

  \maketitle

  \begin{abstract}

   In this study we explore the possibility to use deep learning for the reconstruction of phase images from 4D scanning transmission electron microscopy (4D-STEM) data. The process can be divided into two main steps. First, the complex electron wave function is recovered for a convergent beam electron diffraction pattern (CBED) using a convolutional neural network (CNN). Subsequently a corresponding patch of the phase object is recovered using the phase object approximation (POA). Repeating this for each scan position in a 4D-STEM dataset and combining the patches by complex summation yields the full phase object. Each patch is recovered from a kernel of 3x3 adjacent CBEDs only, which eliminates common, large memory requirements and enables live processing during an experiment. The machine learning pipeline, data generation and the reconstruction algorithm are presented. We demonstrate that the CNN can retrieve phase information beyond the aperture angle, enabling super-resolution imaging. The image contrast formation is evaluated showing a dependence on thickness and atomic column type. Columns containing light and heavy elements can be imaged simultaneously and are distinguishable. The combination of super-resolution, good noise robustness and intuitive image contrast characteristics makes the approach unique among live imaging methods in 4D-STEM.
    \\
    \noindent\textbf{Key Words:} Electron Microscopy, Machine Learning, 4D-STEM, Inverse Problem, Low Dose Imaging, Phase retrieval

  \end{abstract}
\end{frontmatter}

\section*{Introduction}
\label{sec:introduction}

Scanning transmission electron microscopy (STEM) is among the most widely used techniques for the visualization, characterization and quantification of atomic structures in material- and nanoscience. Images are acquired by scanning a sample with an electron probe over a two dimensional (2D) grid. Traditionally, STEM data is collected using annular detectors which measure an integrated intensity over the detector area. This results directly in 2D images where the pixel intensities are proportional to the angular integrated scattered intensity of the electron beam at the corresponding probe position. These images are intuitively interpretable and can further be used for the quantification of atomic structures \citep{VanAert2013}. However, conventional STEM imaging has its limitations, particularly in terms of dose efficiency, resolution and the ability to image light and heavy elements simultaneously. Indeed, in the case of annular dark field (ADF) imaging, the scattering cross section of an atom scales with the atomic number $Z$ roughly $Z^{1.7}$ \citep{Krivanek2010, Yamashita2018}, which often leads to heavy atoms obscuring nearby light elements. On the other hand, imaging very thin or radiation sensitive samples may even be impossible due to the high dose requirements of ADF. This is highly relevant especially for investigations of 2D nano structures and organic compounds. \\
To overcome these limitations much effort has been devoted towards the development of fast pixelated electron detectors which can record an entire convergent beam electron diffraction pattern (CBED) in a reasonable amount of time \citep{Tate2016,Ballabriga2011,Ciston2019,MacLaren2020,haas2021high,jannis2021event}. This enables a set of new imaging modalities, such as ptychography, the calculation of phase contrast \citep{lazic2016phase} and true center of mass (COM) imaging \citep{muller2014atomic}. These imaging modalities can all be considered phase retrieval algorithms in a wider sense \citep{Close2019}, as they all aim to retrieve the projected electrostatic potential of a sample, which directly affects the phase of the transmitted electron wave. Ptychographic methods have been of particular interest for their super-resolution capabilities and the possibility to determine/correct for microscope aberrations as well \citep{WDD_Rodenburg1992,Jiang2018,SSB_RODENBURG1993304}. However, the computational cost and memory requirements for these algorithms are considerable. Iterative algorithms like ePIE \citep{ePIE_MAIDEN20091256, Chen2020} use optimization algorithms to fit an object to a given dataset such that the estimated phase corresponds to the intensity observations across the entire dataset. This is a computationally intensive task and the result is influenced by optimization parameters and convergence criteria. Other non-iterative methods typically use only the bright-field disc for phase reconstructions which imposes limitations on the maximum achievable resolution \citep{SSB_RODENBURG1993304}. They rely on taking Fourier transforms with respect to the probe positions in real space, which means that for conventional Fast Fourier Transform algorithms (FFT), entire datasets (or at least substantial parts of them) have to fit into computer memory, which is becoming increasingly restrictive considering the growing size of 4D-datasets. For these reasons, ptychography has found many useful applications, mainly as a post-experiment data processing analysis step in specialized studies, but has not become a mainstream imaging modality so far. However, there is an increasing interest in using these algorithms interactively during experiments. To that end, live imaging using SSB was recently implemented and demonstrated by \cite{strauch2021live} and \cite{pelz}, as well as live center of mass imaging by \cite{yu2021real}. \\
In the present work we explore the possibility to use machine learning (ML) for dose-efficient phase object (PO) reconstructions with super-resolution in (near) real-time. We show that using a convolutional neural network (CNN) enables fast exit wave retrieval for a given CBED, by using only a 3x3 kernel of adjacent diffraction patterns at a time. The method allows the retrieval of exit waves, with a resolution theoretically only limited by the Nyquist frequency of the detector and thus enables super-resolution imaging at sufficiently high doses. Using only nine CBEDs per probe position in a 4D-STEM dataset implies that the dataset can practically be of arbitrary size and the reconstruction can be performed live during the experiment with appropriate accelerator hardware, such as a modern, single GPU. 
In this paper, the character and capability of the proposed method is discussed in detail and demonstrated on both simulated and experimental data. Comparisons are also made with other possible live processing methods. 

To the best of our knowledge the only reconstruction methods that go beyond utilizing the traditional annular detector and enable live imaging are single sideband ptychography (SSB)\citep{strauch2021live, pelz}, integrated differential phase contrast (iDPC) and integrated center of mass (iCOM) \citep{yu2021real}. This is why we focus our analysis on comparing the results of our proposed method to  those methods.

\section{Materials \& Methods}
\label{sec:materials}

\subsection{Theoretical Framework}
\label{sec:theoretical-framework}

The interaction of fast electrons with thin specimens can be conveniently described with the phase object approximation (POA). As an electron passes through a positive electrostatic potential its wavelength $\lambda$ is temporarily altered which is equivalent to shifting the phase of the electron \citep{kirkland2010}. For cases where the specimen is extremely thin, the propagation of the wave as it goes through the material can be neglected as a reasonable approximation and the real space 3D electrostatic potential of the atomic structure $V_s(\vec{r}, z)$ can be expressed as its integral along the optical axis $z$, resulting in the projected electrostatic potential $v_z(\vec{r}) = \int V_s(\vec{r},z)dz$, which is a function of the vector $\vec{r}$ that spans the remaining two dimensions. With this approximation the exit wave $\psi_{out}(\vec{r})$ is simply the product of the incident wave $\psi_{in}(\vec{r})$ and the object, which can be described by the transmission function $T(\vec{r})$:

\begin{equation}
  \label{eq:POA}
  \psi_{out}(\vec{r}) = \psi_{in}(\vec{r})T(\vec{r})
\end{equation}

\noindent where $T(\vec{r})= \exp\left(i\sigma v_z(\vec{r})\right)$ and $\sigma$ is an interaction parameter (See \citep{kirkland2010} for a more detailed derivation). However, a direct solution of the transmission function according to equation \ref{eq:POA} is only possible if both incident and exit wave are known, while in practice neither of them are known a priori. The incident wave $\psi_{in}(\vec{r})$, can be fairly well approximated as the Fourier transform ($\mathcal{F}$) of the product of the aperture function $A(\vec{k})$ and an aberration-function($\chi(\vec{k})$)-dependent phase shift\citep{Zuo2016}.

\begin{equation}
  \label{eq:psi_in}
  \psi_{in}(\vec{r}) = \mathcal{F}\left[A(\vec{k})\exp{[i\chi(\vec{k})]}\right]
\end{equation}

Here, $\vec{r}$ describes a 2D space at the object plane and $\vec{k}$ describes the reciprocal space. The function $\chi(\vec{k})$, considering only the spherical aberration $C_s$ and defocus $\Delta f$, is given by:

\begin{equation}
  \label{eq:aberration}
  \chi\big(\vec{k}\big) = \pi \lambda k^2 \Big(0.5 C_s \lambda^2 k^2 -\Delta f \Big)
\end{equation}

Assuming that at least the low order aberration parameters of $\chi(\vec{k})$ are known, equations \ref{eq:psi_in} and \ref{eq:aberration} can be used to estimate $\psi_{in}(\vec{k})$. The other piece of missing information is then the exit wave $\psi_{out}(\vec{r})$. From a 4D STEM experiment, only the intensity $|\psi_{out}(\vec{k})|^2$ can be measured (figure \ref{fig:workflow}-\circnum{2}), and thus a method to retrieve the exit waveform on the sample plane based on the information accessible from the experiment is required to solve equation \ref{eq:POA} to get the transmission function $T(\vec{r})$.

Retrieving the phase of $\psi_{out}(\vec{k})$ is a common inverse problem, but is severely complicated in 4D-STEM by the presence of noise to a level that makes even the estimation of $|\psi_{out}(\vec{k})|$ a challenging task in its own.

The idea in this study is to leverage the multislice formalism, incorporating the calculation of electrostatic atomic potentials \citep{Lobato2014} and the frozen phonon approximation \citep{VanDyck2009}, as a forward model to generate a large synthetic dataset. This dataset can then be used to train a convolutional neural network (CNN) to retrieve an estimate of $\psi_{out}$ for any given experiment within the boundaries of the validity of the used forward model and within the parameter space of the training data, which will be discussed in section "\nameref{sec:data_generation}" and is given concisely in table \ref{tab:parameter_space}.

\subsection{General Workflow}
\begin{figure}[htb!]
  \centering
  \includegraphics[width=\linewidth]{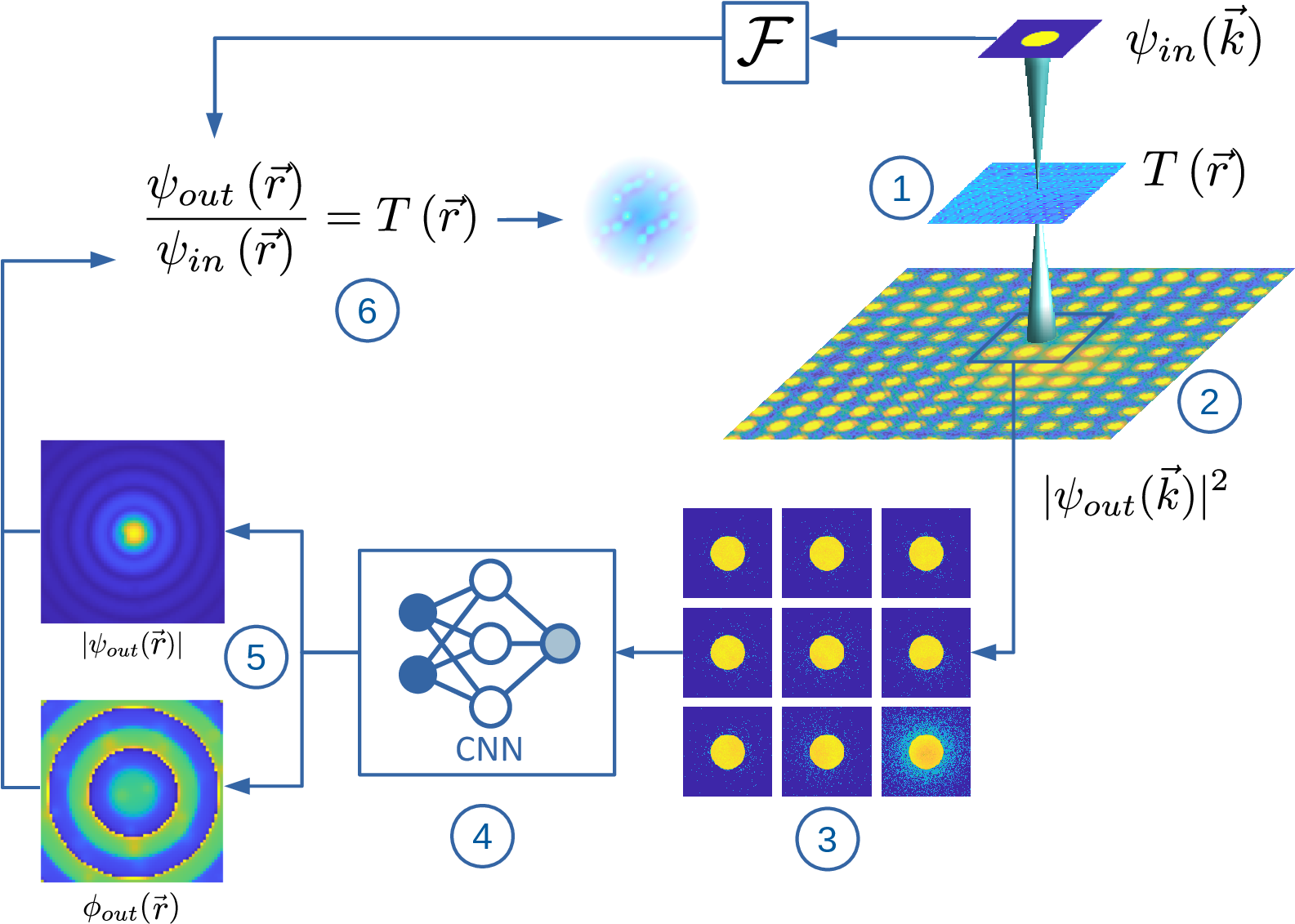}
  \caption{ General workflow: A patch of the phase object \circnum{1} of a 4D-STEM dataset \circnum{2} is reconstructed by extracting a 3x3 kernel of adjacent CBEDs \circnum{3}, using a CNN \circnum{4} to reconstruct the amplitude ($|\psi_{out}(\vec{r})|$) and phase ($\phi_{out}(\vec{r})$) of the exit wave of the central CBED \circnum{5} and using the phase object approximation to reconstruct the object patch \circnum{6} from the reconstructed exit wave and an estimated probe function $\psi_{in}(\vec{r})$. Patches are then stitched together by complex addition to yield a reconstruction of the full phase object.}
  \label{fig:workflow}
\end{figure}

The general concept of the proposed reconstruction method is schematically illustrated in figure \ref{fig:workflow}. The workflow to retrieve the object $T(\vec{r})$ (figure \ref{fig:workflow}-\circnum{1}) of a 4d-STEM dataset (figure \ref{fig:workflow}-\circnum{2}) can be divided into two main steps: firstly, a neural network, trained to solve the inverse problem as outlined in section "\nameref{sec:NN}", reconstructs the phase $\phi_{out}$ and amplitude $|\psi_{out}|$ of the exit wave, based on the intensity measurements of a 3x3 kernel of adjacent diffraction patterns, as depicted in figure \ref{fig:workflow}-(\circnum{3}-\circnum{5}). Secondly, a patch of the object is retrieved from the previously reconstructed exit wave according to equation \ref{eq:POA}, as shown in figure \ref{fig:workflow}-(\circnum{6}). \\

In order to take the relative position of each individual object patch into account, a phase factor is included to the approximated $\psi_{in}(\vec{k})$ and the predicted $\psi_{out}(\vec{k})$ so that their real space counterpart $\psi_{in}(\vec{r})$ and $\psi_{out}(\vec{r})$ are found at the right position. This phase factor $\phi_{shift}$ is a function of the displacement of the probe position in the x and y directions ($\Delta x$ and $\Delta y$) with regard to the center of an array of size $N_x \times N_y$.

\begin{equation}
  \label{eq:shift}
    \phi_{shift}\left(\Delta x, \Delta y\right) = \exp{\left[2 \pi i \left(\frac{\Delta x}{N_x} + \frac{\Delta y}{N_y}\right)\right]}
\end{equation}

The real space wave functions considering the phase factor are:

\begin{equation}
  \label{eq:real_space_wave}
  \begin{split}
    \psi_{in,n}(\vec{r}) &= \mathcal{F} \{\psi_{in}(\vec{k}) \times \phi_{shift,n} \} \\
    \psi_{out,n}(\vec{r}) &= \mathcal{F} \{\psi_{out}(\vec{k}) \times \phi_{shift,n} \}
  \end{split}
\end{equation}
In equation \ref{eq:real_space_wave}, $\mathcal{F}$ is Fourier transformation, and the amount of displacement $\Delta x$ and $\Delta y$ is absorbed into $n$, which specifies the $n^{th}$ probe position.

The phase object approximation assumes that the retrieved object patch should have a homogeneous amplitude of 1 with a phase distribution reflecting the projected potential of the imaged material. However, since the transmitting electron probe carries information mostly from a specific region of the examined sample at the probe position, the retrieved object patches are given a weighting function $\omega_n$ according to the $n^{th}$ incident probe intensity distribution, and the accordingly weighted object patch $\psi_{patch,n}$ is expressed as:

\begin{equation}
  \label{eq:obj_patch}
    T_{patch,n}(\vec{r}) = \frac{\psi_{out,n}(\vec{r})/\psi_{in,n}(\vec{r})}{|\psi_{out,n}(\vec{r})/\psi_{in,n}(\vec{r})|}\times\omega_n(\vec{r})
\end{equation}

with the weighting function $\omega_n$ as:

\begin{equation}
  \label{eq:weight}
  \begin{split}
    &\omega_n(\vec{r}) = \\
    &\left\{ \begin{array}{ccl} \frac{|\psi_{in,n}(\vec{r})|^2}{\sum_{\vec{r}}{|\psi_{in,n}(\vec{r})|^2}} & \mbox{ if } & |\psi_{in,n}(\vec{r})|^2>\frac{1}{10} max(|\psi_{in,n}(\vec{r})|^2) \\ 0 & \mbox{ if } & otherwise \end{array}\right.
  \end{split}
\end{equation}

This procedure is repeated for all real space coordinates in the 4D-STEM dataset and the object patches are combined into the final phase object $T$ by complex addition over $n$ probe positions. 
\begin{equation}
  \label{eq:update}
  T(\vec{r}) = \sum_{n}{T_{patch, n}(\vec{r})}
\end{equation}

The object patch estimations coming from the CNN are not perfect but carry some errors. However, the full object is the combined result of the predictions made at multiple probe positions (equation \ref{eq:update}). Even if one particular prediction is very inaccurate, its impact on the final result is limited as long as it is outweighed by the contributions from neighboring probe positions, which is the case when a significant probe overlap is established by a dense scanning raster. 

\subsection{Training data generation}
\label{sec:data_generation}
\begin{figure*}[htb!]
  \centering
  \includegraphics[width=\linewidth]{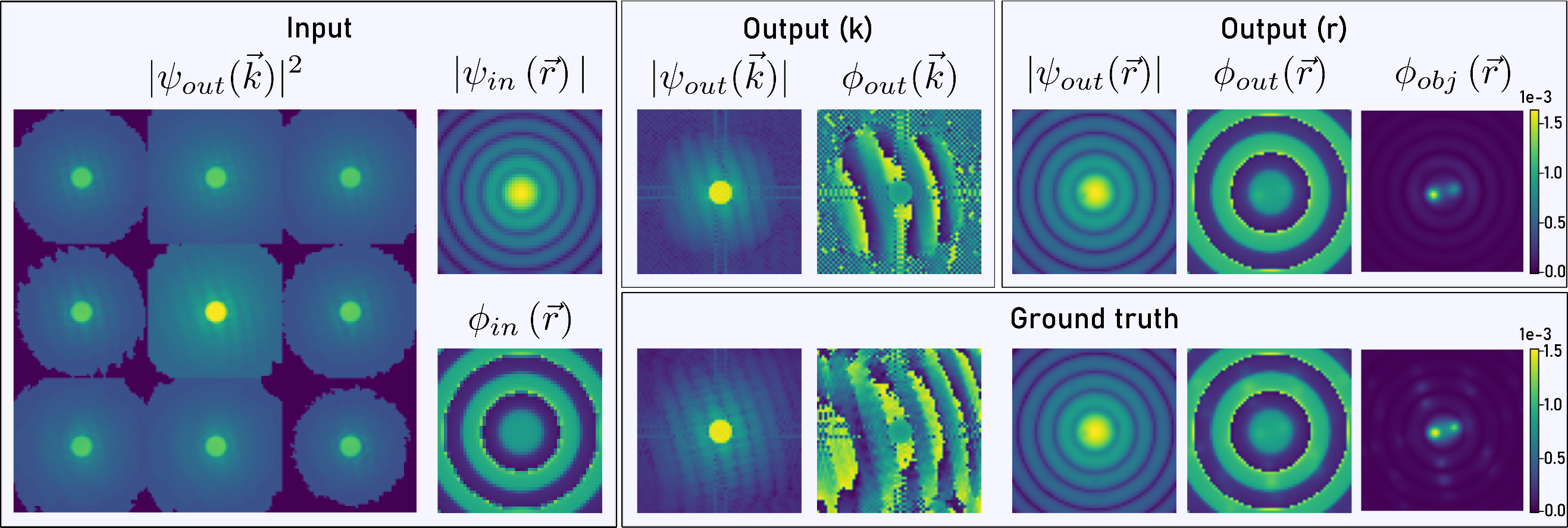}
  \caption{ Example of an exit wave reconstruction taken from the validation dataset, illustrating the inputs and outputs of the CNN, as well as the Fourier transforms of the (real space) exit waves. Intensities and amplitudes are depicted in $\log$ scale.}
  \label{fig:in_out}
\end{figure*}

We created a large synthetic dataset, using atomic structures extracted from \href{https://materialsproject.org/}{the materials project} \citep{Jain2013} database. Based on the unit cell definitions we created bulk specimens in one of the low-index zone axis orientations given in table \ref{tab:parameter_space}. Each sample consists of a 3x3 kernel of simulated CBED patterns as features and the corresponding exit wave in real-space as label. The simulations were performed using the multislice formalism and the frozen phonon approximation. In the implementation given by \citep{Lobato2015,Lobato2016}, following the derivation of \citep{VanDyck2009} the CBED intensity can be expressed as the sum of coherent and incoherent intensities.
\begin{equation}
  \label{eq:exit_wave}
  \Big\langle\big|\psi_{out}\big(\vec{k}\big)\big|^2\Big\rangle = \Big|\big\langle\psi_{out}\big(\vec{k}\big)\big\rangle\Big|^2 + \Big\langle\big|\delta\big(\vec{k},t\big)\big|^2\Big\rangle
\end{equation}
where $\psi_{out}(\vec{k})$ is the exit wave, $\delta$ is a phonon configuration$(t)$-dependent difference and $\langle\rangle$ denotes the average over $t$. This formalism gives us access to the average, coherent, complex wave function. We use this wave function of the central CBED of the 3x3 kernel as labels (i.e. the ground truth training target) and the CBED intensities of all patterns in the kernel, as given in equation \ref{eq:exit_wave} as features (i.e. the CNN input). Only low order aberration parameters $\Delta f$ and $C_{s}$ of $\chi(\vec{k})$ (equation \ref{eq:aberration}) are considered as they are unavoidable and typically have the strongest influence. We assume that including these effects phenomenologically with a constant, small $C_{s}$ and corresponding Scherzer defocus is sufficient. Temporal and spatial incoherence are also not taken into account. This reduces the parameter space considerably but also implies that the method is (for now) limited to aberration corrected, well adjusted microscopes. \\
Further, the dataset includes the CBED-size in $\SI{}{\angstrom}^{-1}$, the aperture size and the acceleration voltage, which allows the computation of the probe function $\psi_{in}(\vec{k})$ within the data pre-processing pipeline during the training using equations \ref{eq:psi_in} and \ref{eq:aberration}. Also the effect of finite electron dose is applied as a data augmentation step during the training, assuming only Poisson noise. To accommodate the possibility that there may be no specimen interacting with the beam, another augmentation step replaces the CBEDs with the probe function intensity with a 3\% chance. 
An example of the resulting training sample inputs and labels is illustrated in figure \ref{fig:in_out} in the "input" and "ground truth" panels respectively. All parameters describing the dataset are summarized in table \ref{tab:parameter_space}. Visualizations of the parameter distributions are shown in supplementary information (figure S.1). The data generation code was published open source under \url{https://github.com/ThFriedrich/ap_data_generation}, as well as the training dataset used \citep{friedrich_thomas_2022_6971200}.

\begin{table}[htb!]
  \begin{tabularx}{\linewidth}{l X}
    \toprule
    Description                     &   Value                     \\
    \midrule
    Acceleration voltage            & $\in$ \{30, 40, 50, 60, 80, 100, 120, 140, 160, 180, 200, 300\}kV   \\
    Step size $(\vec{r})$           & (0.0167...0.33)~\SI{}{\angstrom}     \\
    Convergence angle               & 5...30~mrad                 \\
    Spherical aberration            & 0.001~mm                   \\
    Defocus                         & Scherzer defocus                   \\
    \# Frozen phonons               & 30   \\
    Atom rmsd                       & \SI{0.08}{\angstrom}  \\
    Orientation                     & $\in$ \{[1~1~0], [0~1~1], [1~0~1], [0~0~1], [1~0~0], [0~1~0], [1~1~1]\}    \\
    Thickness                       & <\SI{30}{\angstrom}    \\
    Dose                            & 3...3e9~e/CBED \\
    \# Structures                   & 126,335    \\
    \# Samples                      & 742,688    \\
    \bottomrule
  \end{tabularx}
  \caption{Simulation parameters for the training dataset.}
  \label{tab:parameter_space}
\end{table}

\subsection{Neural Network implementation}
\label{sec:NN}
\noindent
\textbf{CNN architecture} \\
The complexity of the inverse problem practically dictates the use of deep neural networks in this study. The U-Net architecture \citep{UNET} is one of the most popular choices for deep learning applications, since it allows to expand the number of parameters considerably, while maintaining strong backpropagation. Each depth level in the "U-shape" reduces the filter map size which accommodates the retrieval of features of different sizes. This makes the U-NET an extremely versatile and easy to train CNN and hence suitable as a starting point for this project as well \citep{Friedrich2021}. However, since the training target is naturally complex valued, a generic pixel-to-pixel mapping, as commonly employed for CNNs in image processing, cannot be used. Two popular ways of dealing with this exist. Firstly, both, phase and amplitude retrieval problems can essentially be treated separately by defining two outputs and optimizing for dedicated loss functions on the phase and amplitude components of the wave function. This treats the complex wave as two real valued images, which in practice has the advantage that common, highly optimized AI tools can be readily employed. Another approach that naturally lends itself to this kind of problem is using complex valued neural networks \citep{Trabelsi}; an approach that has found relatively few applications so far. However, the theoretical framework for complex CNNs is established \citep{Trabelsi} and implementations have been showing promising results \citep{Virtue, FCU_NET}. We implemented the U-NET architecture for both CNN types to test the main ideas. The complex networks delivered reconstruction results and accuracies of predicted phases, which are, for all intents and purposes, equivalent to the real-valued CNN, while decreasing the speed performance considerably. Since live imaging is an envisioned application, inference speed is a critical concern and the faster real-valued CNN was chosen in the study accordingly.   

The structure of the U-NET is modified to account for physical considerations. The aim of the neural network is to model the electron-specimen interaction. Adding skip connections from the input probe function to the output exit waves (essentially enabling global residual learning) isolates the specimen interaction contributions from the probe function contributions to the exit wave. The CNN therefore does not need to learn to actually model the electron probe. The skip connections have the additional benefit of providing a common template during inference. The training is done on isolated scan points of only 9 CBEDs per specimen, while during inference the probe function should be consistent for the entire dataset, which is a requirement that cannot be captured by any metric during the training. Providing an estimated probe function greatly promotes this consistency during inference. Global residual learning also enhances noise robustness because the probe function serves as template, which is hardly altered if the input is merely noise. 

On the input the dynamic range of the CBEDs are being scaled by taking them to the power of 0.1 in a preprocessing step, which puts more relative weight on the dark field scattered electrons to support exit wave reconstructions beyond the convergence angle. Subsequently each pattern is scaled, according to equation \ref{eq:input_weighting} depending on its distance from the central beam position, where $\zeta_{xy}$ and $\zeta_d$ correspond to the CBED weights adjacent to the central CBED along x and y and on the diagonal, respectively. This is a straightforward way to include the step size $\Delta s$ into the workflow. 
\begin{equation}
\label{eq:input_weighting}
\zeta_{xy} = \frac{1}{\Delta s * 50} \quad \quad
\zeta_d = \frac{1}{\sqrt{2 * \Delta s^2} * 50}
\end{equation}
The constant factor of 50 accounts for the range of step sizes in the training datasets such that all $\zeta$ are between 0 and 1. The effect of this weighting can be seen in the "Input" panel of figure \ref{fig:in_out}, evident by the higher mean intensity of the central CBED.

\begin{figure}[htb!]
  \centering
  \includegraphics[width=\linewidth]{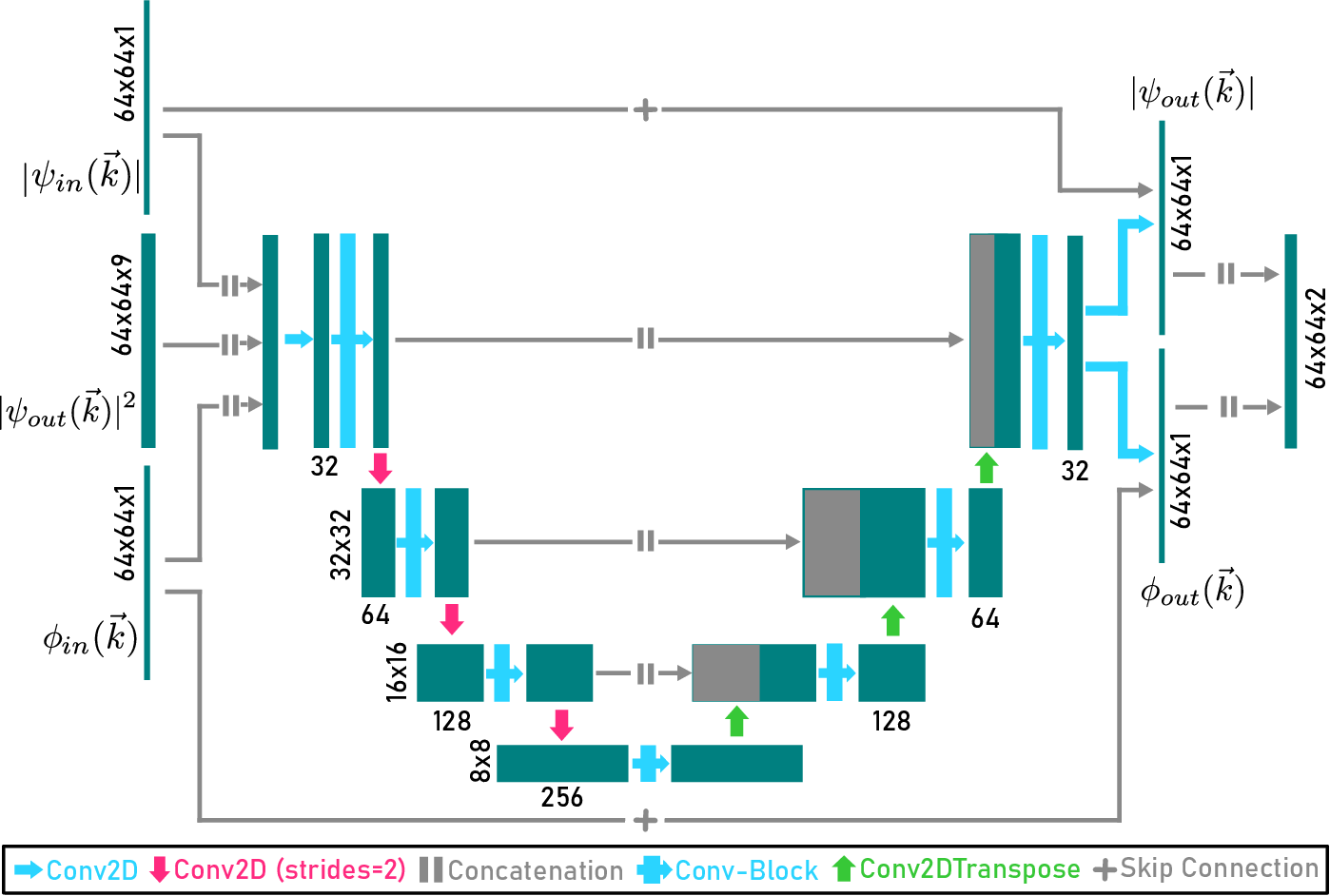}
  \caption{ The CNN architecture used in this study is a modified U-NET with separate, real valued phase and amplitude outputs. The model leverages global residual learning through added skip connections of the probe function to the output. Each "convolution layer" is composed of a 2D convolution layer, batch normalization and a swish activation function. Each "convolution block" consists of 3 consecutive convolution layers. }
  \label{fig:unet}
\end{figure}

Other notable differences as compared to the original U-NET implementation\citep{UNET} are different map sizes, the use of Swish-Activation functions \citep{swish} and the use of strided convolutional layers instead of max-pooling layers to avoid any loss of information when feature map sizes are reduced. Trainable scaling factors on the two output layers for the phase and amplitude were added to scale between the batch-normalized feature maps (with standard deviations of 1 and means of zero) in the CNN and the relatively small target value distributions, which correspond to the difference between exit waves and probe functions. The variables were initialized accordingly with small values of 1e-4 and optimized during training. The amplitude output layer includes a regularization which penalizes integrated exit wave intensities > 1. The phase output layer penalizes values larger than $\pi$ and smaller than $-\pi$. These penalties are added to the loss during training. The resulting overall architecture is depicted in figure \ref{fig:unet}. The tensorflow implementations of the models and individual layers are available open source on Github at \url{https://github.com/ThFriedrich/airpi}. During the training the CNN can process >750,000 sample points in about 6 minutes on a single Nvidia RTX 3080 GPU, indicating that the model itself may be well suited for live processing at rates >2kHz, if the pre- and post-processing pipelines are well optimized.

\noindent
\textbf{Loss function} \\
To facilitate the learning of a general representation for both the phase and amplitude we designed a multi-objective loss function as a linear combination of $\mathcal{L}_2$-losses on the phase and the amplitude in fourier space and in real space. Enforcing the correspondence between $\vec{r}$ and $\vec{k}$-space during the training encourages the CNN to abide to physical constraints. It was also observed that the decomposition of the phase into its $sin$ and $cos$ components facilitates better convergence, compared to just optimizing for the phase directly. This is presumably related to the decompositions being smooth so the CNN does not have to account for phase wrapping effects. Since the probe function is an input to the CNN the object can be solved directly and used in the loss function too. The optimization of the $\mathcal{L}_2$ error of the phase of the object $T(\vec{r})$ directly promotes an agreement of the object with the transmission function, which is practically the most meaningful metric. However, a good quantitative match may be impossible to achieve in certain scenarios (e.g. very low dose). To promote at least a visual match, the object phase is also optimized for its cross correlation $xc$ as given in equation \ref{eq:xc}. Further it was empirically determined that a higher weight on the phase in $(\vec{k})$-space leads to faster convergence and overall better results.
Putting all terms into a sum, for an exit wave $\psi$ with phase $\phi$ and an object phase $\phi_{obj}$ the loss function is given by:
\begin{equation}
  \label{eq:loss}
  \begin{aligned}
    \mathcal{L} & =                                                                                                  \\
                & \mathcal{L}_2(|\psi(\vec{k})|^2)+(\mathcal{L}_2(\sin\phi(\vec{k}))+\mathcal{L}_2(\cos\phi(\vec{k})))*3 \\
    +           & \mathcal{L}_2(|\psi(\vec{r})|^2)+\mathcal{L}_2(\sin\phi(\vec{r}))+\mathcal{L}_2(\cos\phi(\vec{r})) \\  
    +           & \mathcal{L}_2(\phi_{obj}(\vec{r})) + \mathcal{L}_{xc}(\phi_{obj}(\vec{r}))
  \end{aligned}
\end{equation}
with:
\begin{equation}
  \mathcal{L}_2(x)=\left(x_{true}-x_{predicted}\right)^2
\end{equation}

\begin{equation}
  \mathcal{L}_{xc}(x)=\left(1-xc(x_{true},x_{predicted}\right)/2
\end{equation}
where:
\begin{equation}
\label{eq:xc}
xc(x,y)=\frac{\sum\limits_{i,j}\left[(y(i,j)-\overline{y})(x(i,j)-\overline{x})\right]}{\sqrt{\sum\limits_{i,j}[y(i,j)-\overline{y}]^2}\sqrt{\sum\limits_{i,j}[x(i,j)-\overline{x}]^2}} 
\end{equation}
where $x$ and $y$ correspond to pixel values at the locations $(i,j)$ and $\overline{x}$ and $\overline{y}$ are the mean values respectively. \\

\noindent 
\textbf{Training} \\
The training was performed using the Adam optimizer with a learning rate of 5e-4, a batch size of 256 and a momentum setting of $0.9$. The learning rate was decreased by a factor of 0.5 when the validation loss did not decrease for 3 epochs. Convergence was reached after $\approx$50 epochs. After convergence the training was resumed for another 10 epochs with a weighting factor of 10 applied to the $\mathcal{L}_2(\phi_{obj}(\vec{r}))$ term of equation \ref{eq:loss}, which leads to a further small decrease on the object error. This step does not alter the reconstruction results considerably but improves the quantitative match between reconstructed objects and transmission functions somewhat.

\subsection{Experiments \& Simulations}
The demonstrations of the reconstruction methods are performed on both experimental and simulated datasets. For the experiments, probe corrected Thermo Fisher Titan (X-Ant-TEM) and Themis  (Advan-TEM) were used. The former is equipped with a MerlinEM direct electron detector \citep{Ballabriga2011} and the latter with a custom-made Timepix3 detector \citep{poikela2014timepix3}. For the experimental datasets of an \ce{Au} crystal and a \ce{SrTiO3} focused ion beam lamella, which can be found in the the online repository \citep{yu_chu_ping_2021_5572123}, the acceleration voltage is set at 300~kV, the semi convergence angle of the electron beam is 20~mrad, and the scanning step size 0.2~$\text{\si{\angstrom}}$ and 0.185~$\text{\si{\angstrom}}$, respectively. The USY zeolite dataset, which can be found in \citep{jannis_daen_2021_5068510}, is collected at 200~keV, with 12~mrad convergence angle and 0.15~$\text{\si{\angstrom}}$ step size. The simulated twisted bilayer graphene dataset is generated with an acceleration voltage 200~kV, a convergence angle of 25~mrad, and a scan step size 0.2~$\text{\si{\angstrom}}$. The twisted bilayer \ce{MoS2} dataset was simulated with the settings: acceleration voltage 300~kV, convergence angle 20~mrad, and scan step size 0.1~$\text{\si{\angstrom}}$. The \ce{MgO} dataset was created with an acceleration voltage of 300~kV, a convergence angle of 20~mrad, and a scan step size 0.05~$\text{\si{\angstrom}}$. All of the simulated datasets are generated with the MULTEM software \citep{Lobato2015}.

\section{Results \& Discussion}
\label{sec:results}

\subsection{Super-Resolution}
\label{sec:super_resolution}
The reconstruction of the proposed method is based on solving equation \ref{eq:POA} for the object using the incident and the exit wave functions, and therefore the resolution of the method is not explicitly limited by neither the optical conditions of the imaging system nor the sampling density of the electron probe. By reciprocity, the object plane is sampled with a maximum resolution determined by the maximum scattering angle the detector can reach, or the highest angle at which the exit wave can be accurately retrieved, which can potentially result in a higher resolution than permitted by the former two limitations.

\begin{figure}[htb!]
  \centering
  \includegraphics[width=\linewidth]{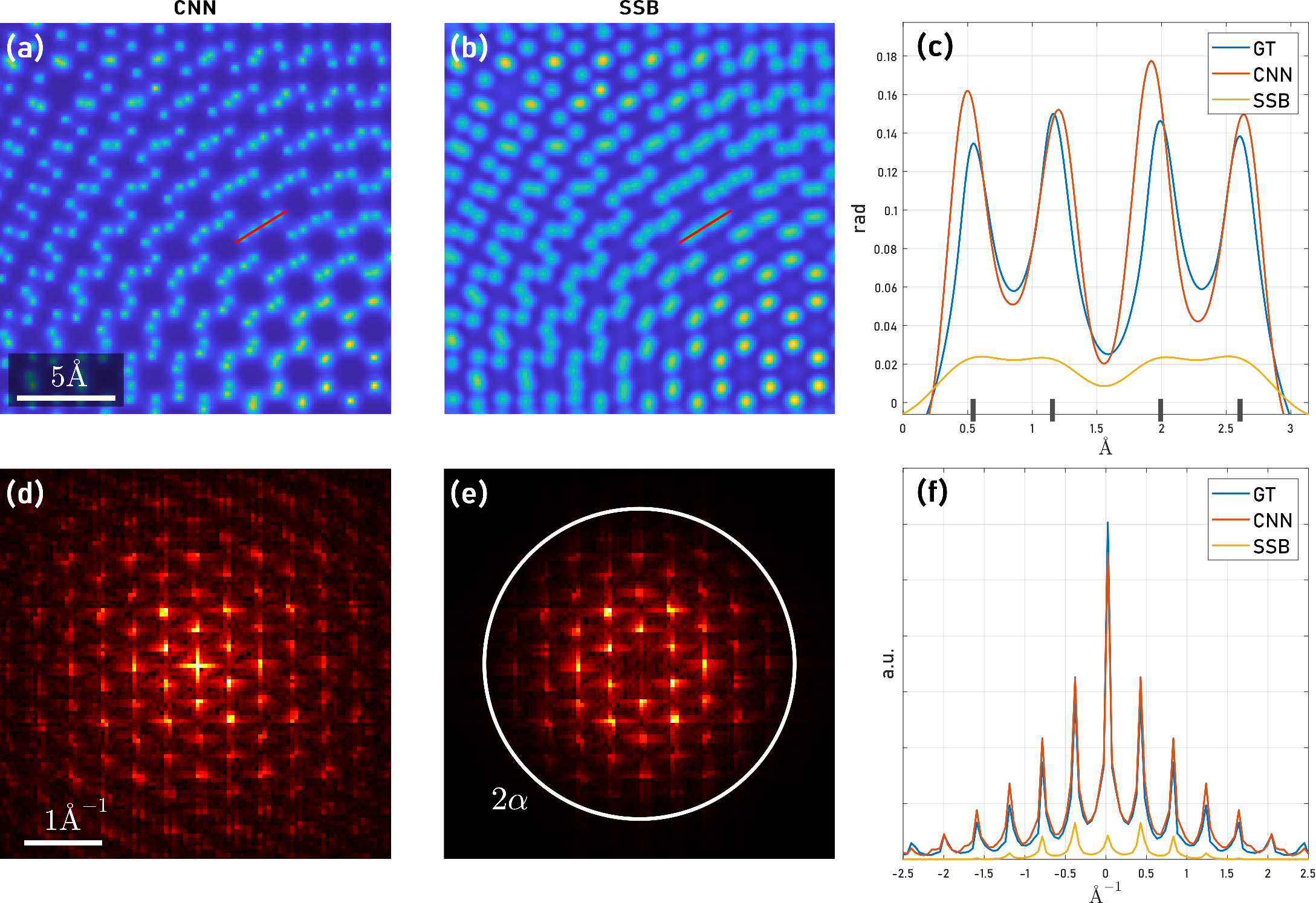}
  \caption{ Demonstration of super resolution capabilities on simulated datasets with infinite dose. Compared are CNN reconstructions and standard SSB ptychography (b). Their corresponding Fourier transformed (FT) intensities show the maximal spatial frequency achieved by each method (d, e). For the FT image of the SSB result, a circle indicating twice the convergence angle is added, which corresponds to the maximal spatial frequency of the method. (c) shows the intensity along the line profile drawn in each image. Markers on the x-axis indicate atom positions. (f) depicts the integrated intensity of the FT images along y-axis.}
  \label{fig:superresolution}
\end{figure}

This super-resolution granted by the knowledge of the exit wave at higher scattering angles is demonstrated by the reconstruction of a simulated dataset of a twisted bilayer graphene sample at infinite dose \citep{friedrich_thomas_2022_7034879}. The result from the CNN reconstruction is shown in figure \ref{fig:superresolution}-a, and compared with a SSB reconstruction in figure \ref{fig:superresolution}-b.

To analyze the spatial frequency achieved by each method, the Fourier transformed (FT) images are presented as well (figure \ref{fig:superresolution}-d, e). The circle in the FT SSB image indicates twice the range of the convergence angle ($\alpha$), which is the upper limit of the spatial frequency of this reconstruction method\citep{SSB_RODENBURG1993304}, and therefore all the frequency components beyond are eliminated. The reconstruction of the CNN successfully retrieves components beyond this limitation, also reflected in the ability to distinguish atoms with very short spacing in between, as can be seen in the atom pairs profile (figure \ref{fig:superresolution}-c)

\begin{figure*}[htb!]
    \centering
    \includegraphics[width=\linewidth]{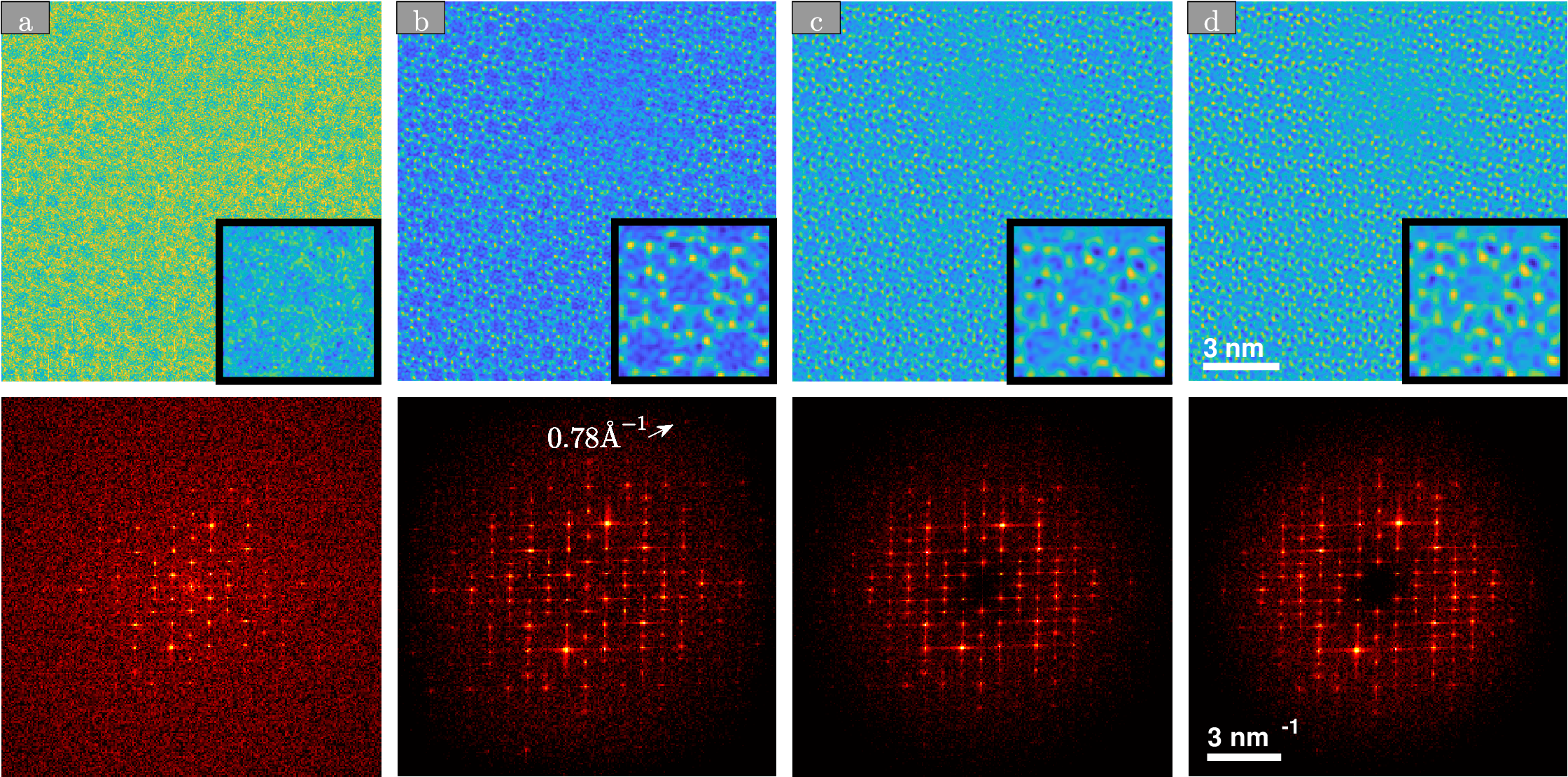}
    \caption{
    Reconstruction results of three different approaches. (a, b) Neural Network performed on dataset without and with data repetition, respectively. (c, d) SSB reconstruction done on datasets without and with data repetition, respectively. The Fourier transforms of the reconstructed images are shown below. Notice that in (a) vertical streaks of very strong intensity can be found, which originate from an unknown defect of the detector, also reported by \citep{jannis2021event}.
    }
    \label{fig:zeolite}
\end{figure*}

The improved resolution capabilities of the method are also demonstrated on an experimental USY-zeolite dataset \citep{jannis_daen_2021_5068510}. To increase the accuracy of the neural network prediction, the CBED at each probe position is replaced by a summation of CBEDs within a $5\times5$ box, while the reconstruction is done with a step size twice as large as the original data. This repetition in data usage increases the effective dose in the dataset, as individual CBED now contains $\frac{25}{4}$-times more electrons and greatly increase the accuracy of the neural network prediction. The actual dose that inflicts damage while interacting with the material, on the other hand, remains the same. For comparison, SSB is performed on the original dataset and a dataset with the same data repetition strategy applied. In figure \ref{fig:zeolite}, it is shown that the last three reconstructions successfully build a clear image of the zeolite crystal structure with atomic level resolution. The CNN reconstruction based on the original dataset does not showcase similar quality, since the dose for individual CBED is too low to make a meaningful prediction of the exit wave, but after data repetition is applied, the neural network gives results that capture details of the material. SSB, on the other hand, does neither benefit nor suffer from the repetition, at least not at a noticeable level. From the Fourier transform of the three images one can estimate the resolution limits of the methods by comparing the most distant frequency component. The neural network reconstruction shows a maximal frequency component at \SI{0.78}{\angstrom^{-1}}, which according to the Raleigh criterion:

\begin{equation}
    \label{eq:raleigh}
    d = \frac{0.61\lambda}{\sin{\alpha}}
\end{equation}

\noindent is equivalent to the resolving power of an un-aberrated perfect optical system of convergence angle 12~mrad, which is the same as the aperture size used in the experiment, at the electron wavelength ($\lambda$) of \SI{0.02}{\angstrom}. As most of the microscopes, even ones equipped with probe corrector, cannot achieve the resolving power given by the Raleigh criterion, the presented method shows the ability to overcome the effect of remaining aberrations, shot noise, and other imperfection in the system to reach a higher resolution.

\subsection{Step Size}
\label{sec:step_size}
\begin{figure}[htb!]
    \centering
    \includegraphics[width=\linewidth]{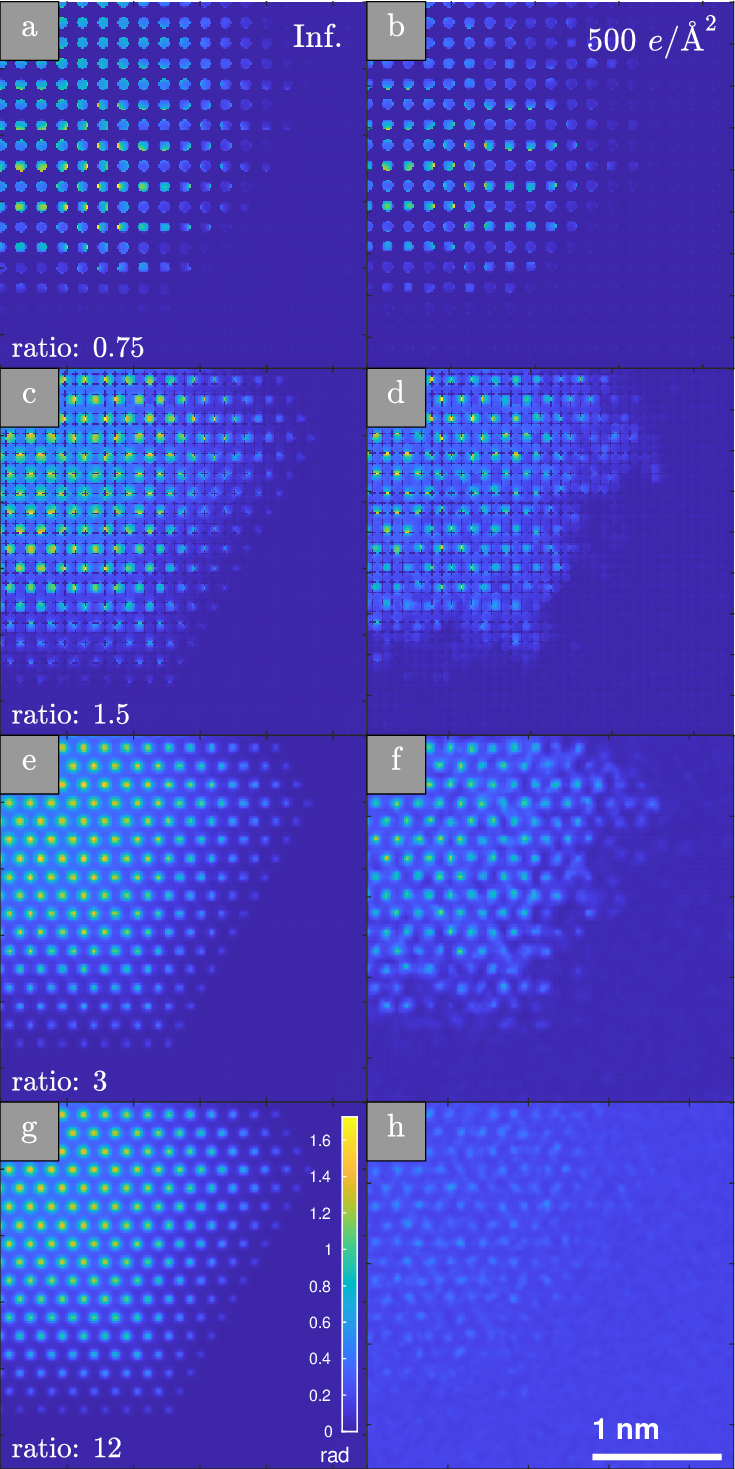}
    \caption{
    Reconstruction results of simulated \ce{MgO} particle. In the left column (a, c, e, g) the images are generated with infinite dose, and in the right (b, d, f, h) the dose is set to be 500 electron per $\si{\angstrom}^2$. Each row of images is constructed with the same step size, as well as the same step/probe-width ratio. The colorbar in the bottom left panel applies to all images in the figure.
    }
    \label{fig:MgO}
\end{figure}

Since the proposed reconstruction method is based on retrieving individual object patches, which are commonly sampled finer than the step the electron beam takes to scan the sample, a rather coarse scanning grid can produce high quality images, as long as a good prediction of the exit wave, and hence the object patch, can be made. This character of the proposed reconstruction method is demonstrated on a simulated \ce{MgO} particle with different step sizes of 0.1, 0.4, 0.8  and 1.6~\si{\angstrom}. The "ratio" values shown in figure \ref{fig:MgO} refer to the ratio of the diameter of the incident probe function ($1.2~\si{\angstrom}$) and the step size, where the probe size for a given convergence angle $\alpha$ in reciprocal \si{\angstrom}ngstrom is defined by the first root of the Bessel function of the first kind and first order: 
\begin{equation}
    \label{eq:bessel}
    d = 2*\frac{3.8317}{\alpha\pi}
\end{equation}
Figures in the left column are generated with infinite dose and therefore the neural network has very detailed knowledge of the amplitude of the exit wave to make accurate predictions. In this case the difference between results of overlap ratio of 12 and overlap ratio of 3 is barely noticeable. By further reducing the scan density, probe positions reach a distance where the weighting function forbids any overlap, as shown in equation \ref{eq:weight}. Despite the weighting function cutoff, which creates blank spaces between the object patches, the actual probe positions used to generate the data overlap with each other just enough, making exit wave predictions possible to maintain the crystal structure to a certain level in the reconstructed image. As the step size reaches $1.6~\si{\angstrom}$ and the ratio drops below 1, the retrieved object patches deteriorate severely and no longer reflect any crystal periodicity. This failure shows that the neural network follows certain physical and mathematical constraints, such as necessary probe overlap for accurate exit wave retrieval, and that it would fail rather then making false predictions that continue to resemble atoms or the crystal. This failure can be identified by the user not only based on the deviation of the resulting image from the expected appearance of the object, but also by the wide blanks left between the object patches, indicating insufficient probe overlap.

The images in the right column of figure \ref{fig:MgO} were generated with the same dose per area. As mentioned in the previous section, the accuracy of the retrieved object patch is not directly related to the total dose in the dataset, but rather to the dose per CBED. By this consideration it follows that larger step sizes work better for the neural network, since this would mean fewer probe positions in the same area and a higher dose at every individual CBED. On the other hand, a certain level of probe overlap is also required for accurate predictions. Therefore, not only the total dose per area, but also the scanning strategy is an important consideration for the proposed method. A balanced scan density will generate better results as compared to a very fine scan grid, even if the total dose per area would be the same. This behavior is illustrated in the images in the right column of figure \ref{fig:MgO}. The noise level is lowered significantly as the step/probe-width ratio drops from 12 to 3 in figures \ref{fig:MgO}-h and \ref{fig:MgO}-f. Ptychographic methods in comparison offer more flexibility in this regard as shown in figures \ref{fig:zeolite}-c and \ref{fig:zeolite}-d.

The noise created by inaccurate predictions also creates different features as the step size changes. As the training is exclusively done on crystalline materials in zone axis orientations, the predicted object patches may show atomic scale features, even if it the input is merely noise. In other words, the frequency transfer function of each object patch is highest at the spatial frequency that would compose an image of an atom. This somewhat dangerous behavior of the neural network is compensated by the stitching of the patches, since atomic-scale features of the noise would not remain sharp as multiple object patches contribute to the same area, and would thus contribute to a cloudy, low intensity background, as seen in figure \ref{fig:MgO}-h. However, when the degree of overlap is reduced, such that the phase value is completely determined by 1 or few object patches, the risk to observe a false, atom-like feature, such as the ones seen in figures \ref{fig:MgO}-f and \ref{fig:MgO}-d, greatly increases. Awareness of this effect is therefore important when using the method and large step-size scanning patterns should be treated with extra caution. The other effect of noise is that the phase value drops in cases of very few electrons per CBED. This is essentially related to the failure of the CNN at making accurate phase object predictions, but it also shows that as long as sufficient dose is present in the $3\times3$ CBED input set, the phase value retrieved is not strongly related to step size. 

\subsection{Contrast Analysis}

\begin{figure}[htb!]
    \centering
    \includegraphics[width=\linewidth]{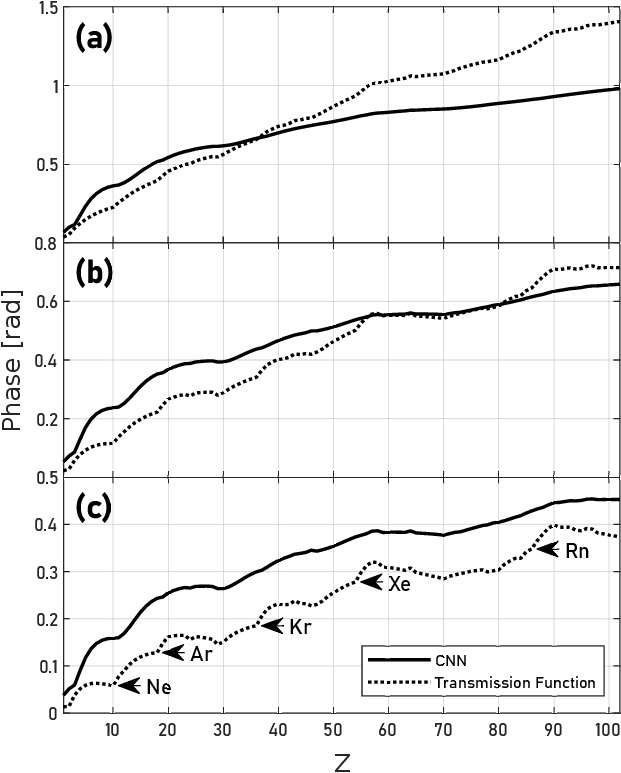}
    \caption{Phase response of the CNN compared to ground truth transmission functions for simulations of single atoms throughout the periodic table at infinite dose for (a) peak intensity, (b) mean over 3x3 pixels around the atomic position, and (c)  mean over 5x5 pixels around the atomic position.
    }
    \label{fig:periodic_table}
\end{figure}

As outlined in section "\nameref{sec:theoretical-framework}", the phase of the object is proportional to the electrostatic potential. By this relation atomic species should, at least within the boundaries of the phase object approximation, be distinguishable. To verify whether this requirement holds true for the CNN reconstructions, 4D STEM datasets of isolated, single atoms for each species in the periodic table up to Z=103 are simulated individually with a step size of \SI{0.2}{\angstrom} and a simulation box size of 3$\times$\SI{3}{\angstrom}. The retrieved phase objects of these datasets are compared to the ground truth transmission function, which is based on the parameterization by \citep{lobato2014accurate}. The comparison is carried out by taking averaged phase values from pixels within various ranges. In figure \ref{fig:periodic_table}, from top to bottom, the curves show phase values at the peak only (figure \ref{fig:periodic_table}-a), averaged phase over $3\times3$ pixels around the atomic position (figure \ref{fig:periodic_table}-b), and the average over $5\times5$ pixels (\ref{fig:periodic_table}-c). Both, the curves of the ground truth transmission function and the CNN predictions, generally increase against atomic number, with the exception of certain dips at larger averaging ranges. This effect stems from different electron orbital distributions in the radial direction, and thus only the phase value averages at larger ranges are sensitive to this difference. The CNN predictions obviously also preserve these sub-atomic level details to some extent, as the shape of the curves bear strong resemblance to the ground truth curves. 
Although the reconstructed phase values and the transmission functions do not match exactly, the predictions are accurate enough, such that the phase values of the reconstructed objects are indeed useful as an indicator for different atomic species, potentially even allowing semi-quantitatively predicting the exact atomic species.

\begin{figure}[htb!]
  \centering
  \includegraphics[width=0.9\linewidth]{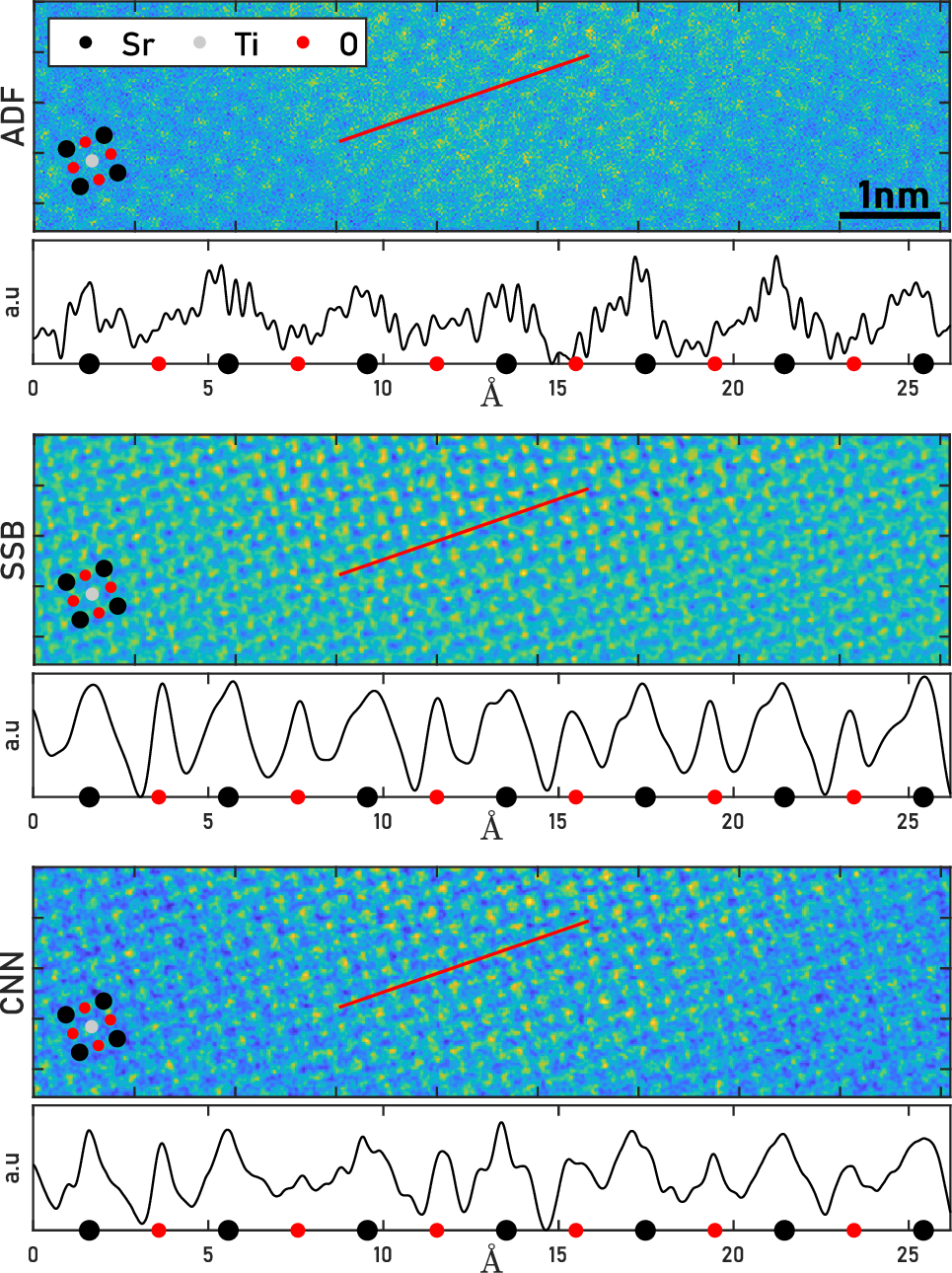}
  \caption{ Reconstructed images of \ce{SrTiO3} with (a) virtual ADF detector, (b) SSB, and (c) CNN. Line profiles are drawn to illustrate contrast between the heavier \ce{Sr} columns and the lighter \ce{O} columns, which are also indicated respectively with dark blue and red arrows in the image.}
  \label{fig:sto}
\end{figure}

For thick samples the method is not expected to yield results in quantitative agreement with projected potentials, because even if the neural network would retrieve the correct exit wave, the reconstruction algorithm is still based on the POA and inherits its limitations. The analysis of thicker samples presented here is therefore done in a more qualitative/empirical manner and comparisons are made against ADF imaging and SSB. ADF images are well known for their strong contrast related to the scattering power of the imaged object, and thus are suitable for examining the thickness variation of the sample \citep{de2016statstem} and local elemental compositions \citep{pennycook1988chemically}. The contrast of SSB reconstructions is not as strong as ADF \citep{yang2016simultaneous}, yet the method is often used to study crystals containing elements of a wide range of atomic numbers due to its ability to image heavy and light atomic columns at the same time with distinguishable contrast \citep{lozano2018low}.
Albeit a quantitative match can hardly be expected, it is important to verify whether reconstructed phase images still reflect the relative projected potentials of thicker samples to avoid misinterpretations. To that end an experimental dataset of a \ce{SrTiO3} FIB-lamella was analyzed. The reconstruction results are presented and compared to ADF and SSB images in figure \ref{fig:sto}. 

For ADF imaging the contrast difference between the \ce{Sr} and \ce{O} columns is too large, making it difficult to locate the \ce{O} columns without the help of the profiling. On the other hand, the SSB reconstruction does successfully image both atomic column types. While the peak intensities of the columns are ambiguous, they can still be distinguished by their corresponding size. The \ce{O} columns are sharper than the \ce{Sr} ones, indicating that an integrated signal from the area of each column could still be used as a reference of the local projected potential. 
The CNN reconstruction exhibits the advantage of both: while the light atom columns are observable, both the intensity and size differences are large enough to distinguish their types. This confirms, that the Z-contrast sensitivity is preserved for thicker specimen.

\begin{figure*}[htb!]
  \centering
  \includegraphics[width=0.9\linewidth]{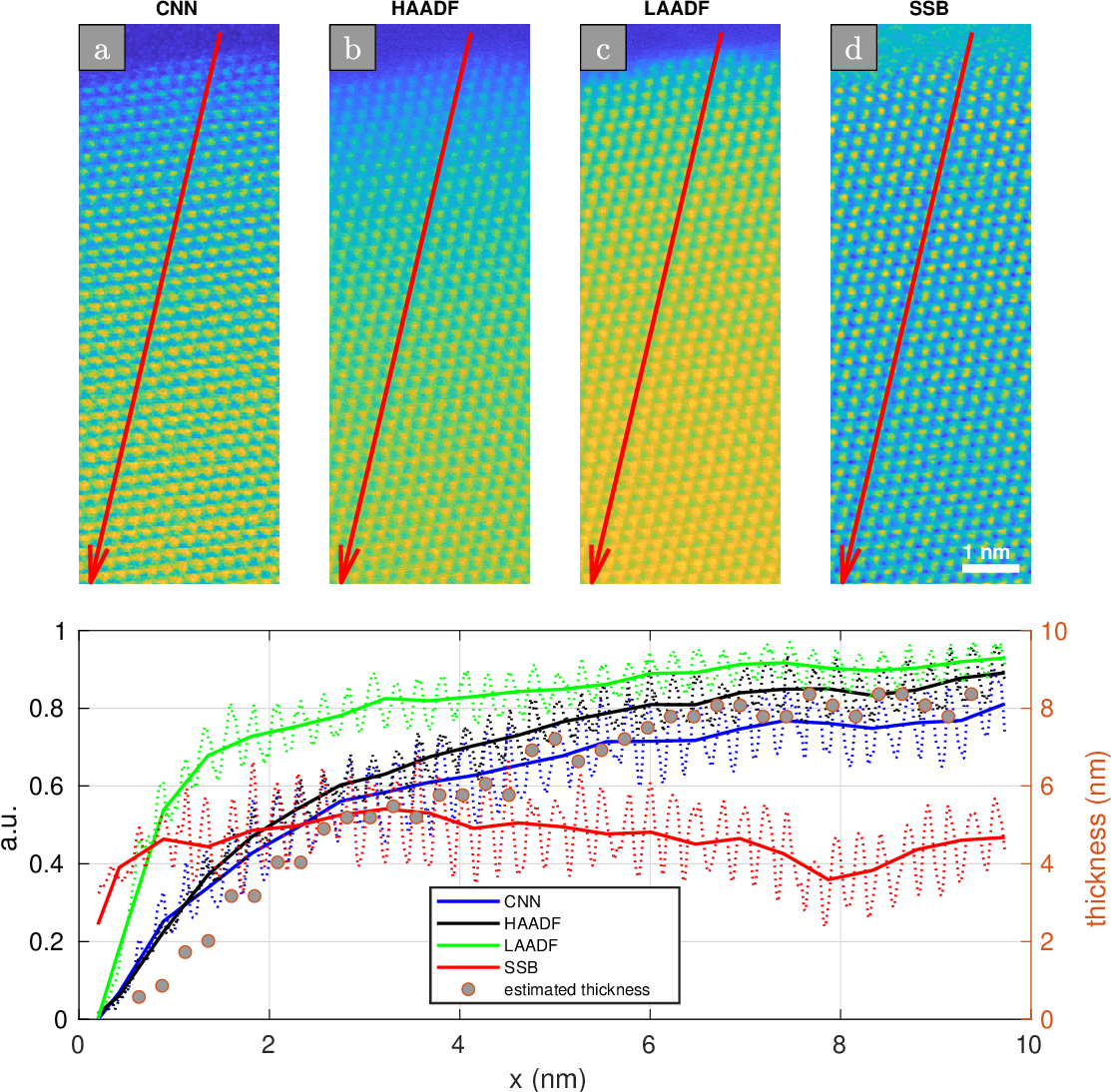}
  \caption{Reconstructed images of an edge of a Au crystal using f.l.t.r: CNN reconstruction, HAADF, LAADF and SSB. Line profiles across the nanorod illustrate the thickness dependence of the corresponding signals.}
  \label{fig:thickness}
\end{figure*}

To further investigate the thickness dependence of the retrieved phase signal, an experimental dataset of the tip of an \ce{Au} nanorod was used. As shown in figure \ref{fig:thickness}, the intensity recorded by the virtual low angle ADF (LAADF) detector (20~mrad to 30~mrad) and HAADF detector (45~mrad and beyond) increases from the top to the bottom of the image. Based on statistical analysis of the HAADF signal to retrieve atom counts in each column \citep{de2016statstem}, the thickest part in the image is about 9~nm. A line profile is then drawn for each imaging method presented in the figure. For the two ADF imaging methods at different collection angle, the profiles show monotonic increase at different pace against thickness, while the SSB profile only shows locations but the intensity is not correlated to the thickness of the atomic column. Compared to these profiles, the CNN reconstruction appears to be qualitatively most similar to the one of HAADF, and correlates with the estimated thickness accordingly. It should be noted that the maximum thickness of $\approx$9~nm is well outside the parameter range of the training data (<3~nm). Also the \ce{SrTiO3}-sample, being a FIB-lamella, should be well beyond 3~nm thickness. This means that these examples also demonstrate extrapolation capabilities of the CNN, which albeit being quantitatively arguably inaccurate, may still provide very useful reconstructions for imaging purposes. The strong resemblance with HAADF images at larger thicknesses may in fact be a very desirable characteristic, as it aligns well with many microscopists' experience and intuition.

By comparing the reconstructed images from the \ce{Au} crystal and \ce{SrTiO3}, one would notice that SSB recovers contrast of higher spatial frequencies, such as the intensity and shape of atomic columns, but it does not recover long range features induced by e.g. thickness variation. This is due to the band-pass nature of the method \citep{yang2015efficient, o2021contrast}. Long range features are build with low frequency components, and thus for reconstruction methods that filter out, or cannot utilize signals that fall in the low frequency end, these features are lost. For the CNN reconstruction, the object patches are also localized and no information beyond one probe position away are shared among the prediction of the exit wave. Therefore, the existence of long range contrast variation can only be attributed to a good prediction accuracy of the CNN.

\subsection{Noise robustness}
\begin{figure}[htb!]
  \centering
  \includegraphics[width=\linewidth]{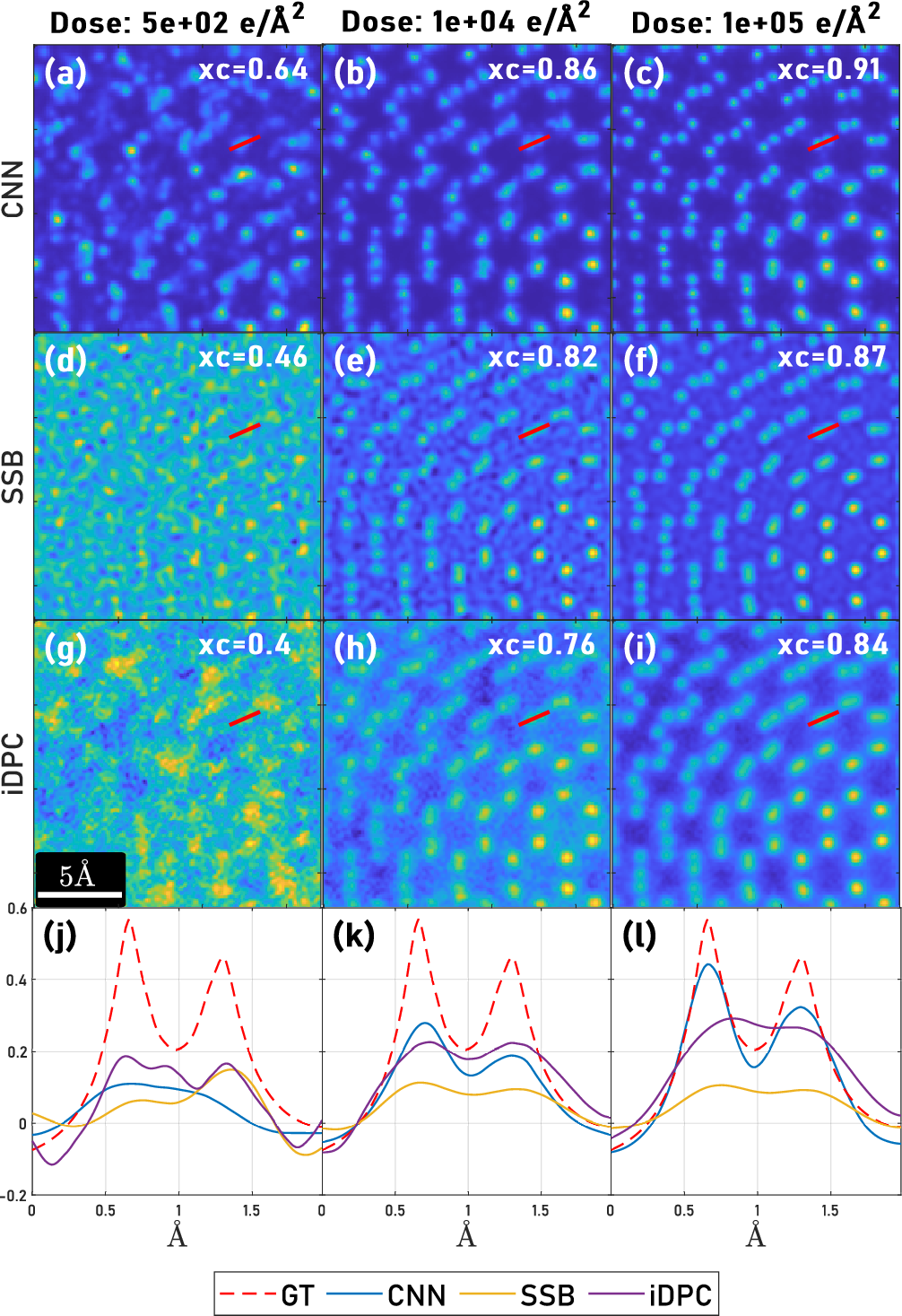}
  \caption{ The dose dependency of the proposed reconstruction method is demonstrated for a simulated dataset of twisted bilayer \ce{MoS2} (top row panels) and compared to the corresponding SSB (2nd row) and iDPC(3rd row) reconstructions. Cross correlation values $xc$ are given with respect to the ground truth phase object in panel of each result image. In (j, k, l) line profiles averaged in the perpendicular direction over \SI{1.6}{\angstrom} are drawn accross a $Mo - S_2$ pair in all the images, and shown with the ground truth. iDPC values were normalized by the maximum value of the transmission function.}
  \label{fig:doses}
\end{figure}

The performance of the method under different dose conditions is demonstrated and analyzed on a simulated dataset of a twisted \ce{MoS2} bilayer. The dose used for the reconstruction ranges from $500$ to $10^5 e / \si{\angstrom}^2$, and the dataset is processed by the proposed method, SSB, and iDPC. The methods reconstructions are illustrated in figure \ref{fig:doses} and compared against the ground truth transmission function in terms of their normalized cross correlations (equation \ref{eq:xc}).

It is evident that the method is strong in all three dose conditions, as can be seen visually in figure \ref{fig:doses}, and is confirmed by a cross correlations higher than those of the corresponding SSB and iDPC reconstructions. The higher cross correlation is not only due to a lower noise level, but also the generally better matched atom shape and phase value with respect to the ground truth, as indicated in the line profiles drawn under each dose condition (figures \ref{fig:doses}j-l). The sharper atom shape is due to the superior resolution, which is confirmed by the better distinction of close atom pairs, as well as by the higher frequency component found in the Fourier transform of the image (figure S2). In figure \ref{fig:doses}-l, the atom profile of the CNN reconstruction at high dose almost strikes a perfect match with the transmission function. Note that the CNN- and SSB-lines are not normalized or shifted, indicating that a very accurate wave retrieval is achieved by the CNN. Additionally, the contrast of the CNN results allows to distinguish \ce{Mo} and \ce{S2}, which is more difficult with other methods, as their difference is much smaller. At low dose, the line profiles suggest a stronger low-dose robustness may be found in iDPC, as the two peaks are preserved in the reconstructed image. However, the signal almost completely falls into the noise level, as confirmed by the Fourier transform, and thus this seemingly better low-dose performance could very likely be a coincidental noise distribution. The given example highlights the potential of the proposed method for low dose imaging. As pointed out in sections "\nameref{sec:super_resolution}" and \nameref{sec:step_size}" and illustrated in figures \ref{fig:zeolite} and \ref{fig:MgO} respectively, the noise robustness may depend substantially on the step size and the effective dose per CBED. To gain an advantage over other methods in this regard this context needs to be taken into account and scanning strategies adapted accordingly.

\section{Conclusion}
\label{sec:summary}
This paper presents a new computational imaging method, leveraging a CNN to retrieve complex exit wave functions from CBEDs and an algorithm to reconstruct the phase object from the predictions of the neural network. Since the exit waves are retrieved for each real-space coordinate in a 4D-STEM dataset, based only on a small kernel of adjacent diffraction patterns, the method can be employed in a sequential manner, thus enabling live imaging during an experiment. 
The machine learning  system is based on a well established model but streamlined to the task at hand and adapted to account for physical constraints and considerations. The model was trained on a large synthetic dataset of multislice simulations.  Large and higher order aberrations, as well as CBED distortions, like non-centricity, geometric distortions and hot/dead pixels are not considered in the training data. Therefore, experimental data may require a pre-processing step. The range of practical conditions for which the method works reliably is therefore arguably limited accordingly to aberration corrected, well adjusted instruments at this stage. The trained model, code and training data are publicly available as summarized in the supplementary information. 
In the discussion multiple unique characteristics and advantages of the method are demonstrated. The CNN-based reconstruction is shown to enable higher resolutions than any other live-imaging-capable method considered, on simulated, as well as on experimental data, provided that a sufficiently high dose-per-CBED is maintained. 
In correspondence to this consideration the effect of the step size is analyzed. While a better estimation of the exit wave is obtained with the electron dose-per-area distributed across fewer probe positions, some probe overlap is necessary to insure the accuracy of the exit wave retrieval. Hence, the method is most suitably applied at a balanced scan density. If these considerations are taken into account the reconstruction method can be very dose efficient.

The Z-contrast was analyzed on single atom-simulations across the periodic table. The phase signal of the reconstructions could indeed be linked qualitatively to atomic properties and a semi-quantitative analysis of thin specimen within the limits of the POA, was shown to be possible.
We confirmed the contrast sensitivity to atomic species and sample thickness on experimental datasets of a \ce{SrTiO3} FIB-lamella and an \ce{Au} nanorod respectively. The observed monotonic increase of the phase signal with thickness and nearly monotonic increase with atomic number indicates that quantitative analyses based on the reconstruction results may be feasible.

Generally, we believe the proposed method presents an attractive imaging modality for its super-resolution capability, high noise robustness, and the feasibility of qualitative or even quantitative contrast analysis. While further studies would be necessary to obtain a more detailed view on the model performance over the entire parameter space (and beyond), we could already show that the method is robust for a wide range of practically meaningful applications, even exhibiting reasonably good extrapolation behavior well beyond the maximum sample thickness of the training data. The fact that none of the examples shown in this study exist in those exact configurations in the training data, further indicates that the system generalizes well within the parameter interpolation range as well. \\

\noindent\small\color{Maroon}\textbf{Data availability }\color{Black}
The training data, the trained model, all implementations, scripts and the data generation codes are publicly available under their respective license terms as summarized in the supplementary information.\\

\noindent\small\color{Maroon}\textbf{Competing interests }\color{Black}
The authors declare no competing interests.\\

\noindent\small\color{Maroon}\textbf{Acknowledgements }\color{Black}
We acknowledge funding from the European Research Council (ERC) under the European Union’s Horizon 2020 research and innovation program (Grant Agreement No. 770887 PICOMETRICS) and funding from the European Union’s Horizon 2020 research and innovation program under grant agreement No. 823717 ESTEEM3. J.V. and S.V.A acknowledge funding from the University of Antwerp through a TOP BOF project. The direct electron detector (Merlin, Medipix3, Quantum Detectors) was funded by the Hercules fund from the Flemish Government. This work was supported by the FWO and FNRS within the 2Dto3D project of the EOS program (grant number 30489208).

\normalsize



\onecolumn


\appendix
\renewcommand\thefigure{S\arabic{figure}} 
\setcounter{figure}{0}   
\renewcommand\thetable{S\arabic{table}} 
\setcounter{table}{0}  

\section*{Supporting Information}

\subsection*{Code and Data}

\noindent \textbf{Trained Neural Network and Reconstruction implementations:}
\begin{tabbing}
authors: \=  Thomas Friedrich and Chu-Ping Yu \\
title: \> airpi \\
license: \> GNU GPL3 \\
url: \> \href{https://github.com/ThFriedrich/airpi}{https://github.com/ThFriedrich/airpi} 
\end{tabbing}

\noindent \textbf{Training Datsets:}
\begin{tabbing}
authors: \=  Thomas Friedrich, Chu-Ping Yu, Jo Verbeeck and Sandra van Aert \\
title: \> Phase Object Reconstruction for 4D-STEM using Deep Learning, (4D-STEM Training Data) \\
doi: \> 10.5281/zenodo.6971200 \\
license: \> Creative Commons Attribution 4.0 International Public License \\
url: \> \href{https://zenodo.org/record/6971200#.YwflDmxBxhE}{https://zenodo.org/record/6971200\#.YwflDmxBxhE}\\
\end{tabbing}

\noindent \textbf{Data generation code:}
\begin{tabbing}
authors: \=  Thomas Friedrich and Chu-Ping Yu \\
title: \> ap\textunderscore data\textunderscore generation \\
license: \> GNU GPL3 \\
url: \> \href{https://github.com/ThFriedrich/ap_data_generation}{https://github.com/ThFriedrich/ap\textunderscore data\textunderscore generation} 
\end{tabbing}

\noindent \textbf{STO datset:} 
\begin{tabbing}
authors: \=  Chu-Ping Yu, Thomas Friedrich, Daen Jannis, Xie Xiaobin, Sandra van Aert and Jo Verbeeck  \\
title: \> Real Time Integration Center of Mass (riCOM) Reconstruction for 4D-STEM \\
doi: \> 10.5281/zenodo.6971200 \\
license: \> Creative Commons Attribution 4.0 International Public License \\
url: \> \href{https://doi.org/10.5281/zenodo.5572123}{https://doi.org/10.5281/zenodo.5572123}\\
\end{tabbing}

\noindent \textbf{Zeolite datset:} 
\begin{tabbing}
authors: \=  Daen Jannis, Christoph Hofer, Chuang Gao, Xiaobin Xie, Armand Béché, \\ 
Timothy  J. Pennycook and Jo Verbeeck  \\
title: \> Event driven 4D STEM acquisition with a Timepix3 detector: microsecond dwelltime  and faster scans \\ 
for high precision and low dose applications \\
doi: \> 10.5281/zenodo.6971200 \\
license: \> Creative Commons Attribution 4.0 International Public License \\
url: \> \href{https://doi.org/10.5281/zenodo.5068510}{https://doi.org/10.5281/zenodo.5068510}\\
\end{tabbing}

\noindent \textbf{Example Datsets:}
\begin{tabbing}
authors: \=  Thomas Friedrich, Chu-Ping Yu, Jo Verbeeck and Sandra van Aert \\
title: \> Phase Object Reconstruction for 4D-STEM using Deep Learning, (4D-STEM Examjple Data) \\
doi: \> 10.5281/zenodo.7034879 \\
license: \> Creative Commons Attribution 4.0 International Public License \\
url: \> \href{https://doi.org/10.5281/zenodo.7034879}{https://doi.org/10.5281/zenodo.7034879}\\
\end{tabbing}

\newpage
\subsection*{Additional data and figures}

\begin{figure}[htb]
  \centering
  \includegraphics[width=\linewidth]{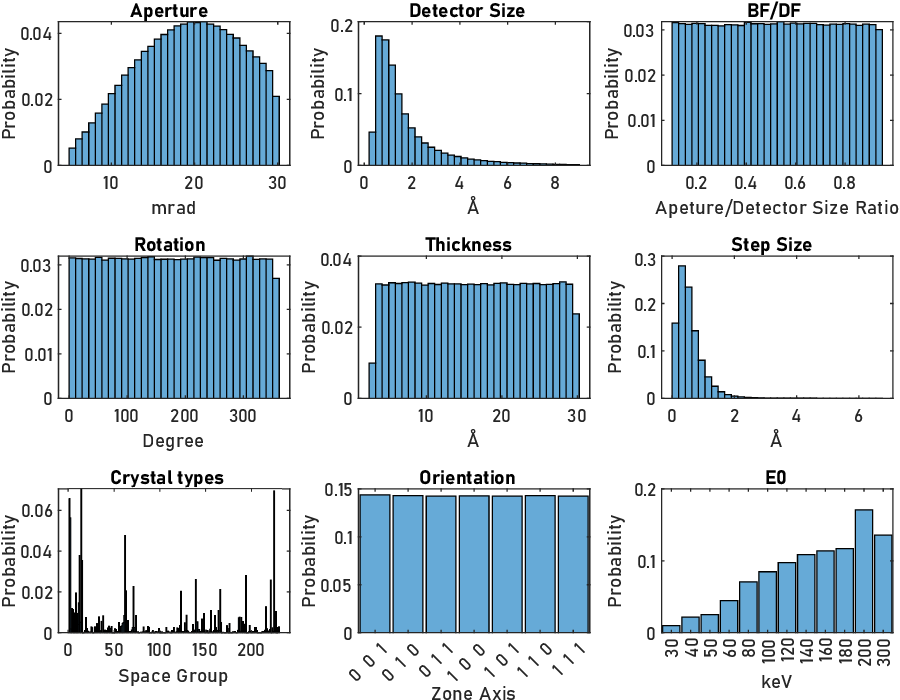}
  \caption{ Parameter distributions of most important simulation parameters of the trainnig dataset, consisting of 742,688 samples.}
  \label{fig:data_distribution}
\end{figure}

\begin{table}[!ht]
    \raggedleft
    \caption{Table summarizing all occurrences of material structures used in this paper in the training data with corresponding simulation settings.}
    \begin{tabularx}{\linewidth}{l l l l l l l l l l}
    \toprule
        mp\_id & Composition & SG & Zone Axis & Rotation & Thickness & E0 & Aperture & Detector Size & $\Delta s$ \\ 
        \midrule
        Gold \\
        mp-81 & Au4 & 225 & 1  1  1 & 338.3 & 15.8 & 300 & 20.94 & 5.216 & 0.180 \\ 
        mp-81 & Au4 & 225 & 0  0  1 & 50.7 & 22.8 & 200 & 8.02 & 0.522 & 1.357 \\ 
        mp-81 & Au4 & 225 & 1  0  0 & 84.7 & 9.1 & 100 & 22.63 & 0.955 & 0.617 \\ 
        mp-81 & Au4 & 225 & 0  1  1 & 148.0 & 18.5 & 180 & 24.97 & 3.073 & 0.375 \\ 
        \midrule
        Graphene \\
        mp-48 & C4 & 194 & 1  0  1 & 2.6 & 14.1 & 80 & 15.63 & 0.398 & 0.364 \\ 
        mp-48 & C4 & 194 & 0  1  0 & 273.4 & 13.0 & 160 & 25.28 & 4.242 & 0.405 \\ 
        mp-48 & C4 & 194 & 1  0  0 & 217.1 & 7.6 & 300 & 20.63 & 1.871 & 0.302 \\ 
        mp-48 & C4 & 194 & 0  1  0 & 168.9 & 24.1 & 30 & 28.43 & 0.473 & 1.070 \\ 
        \midrule
        \ce{MgO} \\
        mp-1265 & Mg4 O4 & 225 & 1  1  0 & 274.9 & 6.2 & 200 & 22.50 & 1.031 & 0.348 \\ 
        \midrule
        \ce{MoS2} \\
        mp-2815 & Mo2 S4 & 194 & 1  0  1 & 75.8 & 8.1 & 140 & 10.60 & 1.355 & 0.381 \\ 
        mp-2815 & Mo2 S4 & 194 & 0  1  1 & 45.9 & 7.9 & 120 & 26.46 & 0.855 & 0.685 \\ 
        mp-2815 & Mo2 S4 & 194 & 0  1  1 & 330.8 & 23.1 & 200 & 29.22 & 1.370 & 0.145 \\ 
        mp-2815 & Mo2 S4 & 194 & 1  0  1 & 256.0 & 9.4 & 180 & 18.86 & 1.945 & 0.191 \\ 
        mp-2815 & Mo2 S4 & 194 & 0  1  1 & 22.1 & 29.7 & 160 & 20.48 & 0.760 & 0.471 \\ 
        \midrule
        \ce{SrTiO3} \\
        mp-5229 & Sr1 Ti1 O3 & 221 & 1  1  0 & 300.6 & 14.2 & 120 & 20.42 & 3.693 & 0.755 \\ 
        mp-5229 & Sr1 Ti1 O3 & 221 & 0  1  1 & 64.8 & 14.7 & 100 & 5.57 & 0.581 & 1.416 \\ 
        mp-5229 & Sr1 Ti1 O3 & 221 & 0  1  0 & 356.4 & 12.1 & 300 & 25.83 & 1.796 & 0.383 \\ 
        mp-5229 & Sr1 Ti1 O3 & 221 & 1  0  0 & 122.6 & 28.0 & 160 & 22.50 & 0.845 & 0.223 \\ 
        mp-5229 & Sr1 Ti1 O3 & 221 & 1  0  1 & 346.2 & 15.4 & 100 & 7.88 & 0.461 & 0.892 \\ 
        mp-5229 & Sr1 Ti1 O3 & 221 & 0  1  0 & 194.9 & 9.1 & 40 & 22.55 & 0.486 & 0.426 \\ 
        mp-5229 & Sr1 Ti1 O3 & 221 & 0  1  0 & 1.8 & 23.5 & 300 & 20.42 & 1.187 & 0.363 \\ 
        mp-5229 & Sr1 Ti1 O3 & 221 & 0  1  1 & 139.2 & 15.3 & 160 & 22.61 & 1.072 & 0.446 \\ 
        \bottomrule
    \end{tabularx}
\end{table}

\begin{figure}[htb]
  \centering
  \includegraphics[width=0.7\linewidth]{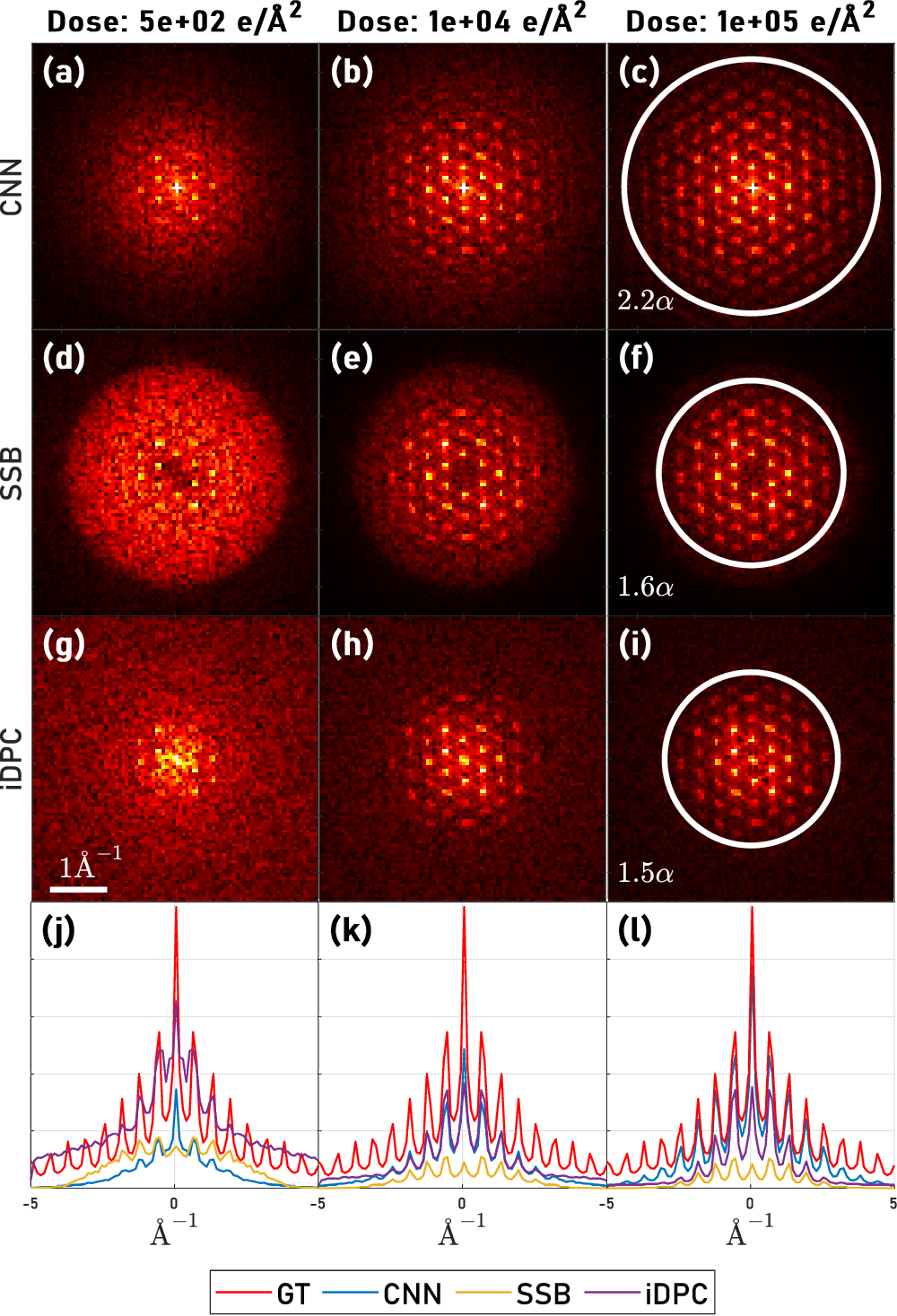}
  \caption{Fourier transforms of the the the MoS2 dataset presented in the paper.}
  \label{fig:ffts_mos2}
\end{figure}

\begin{figure}[htb]
  \centering
  \includegraphics[width=\linewidth]{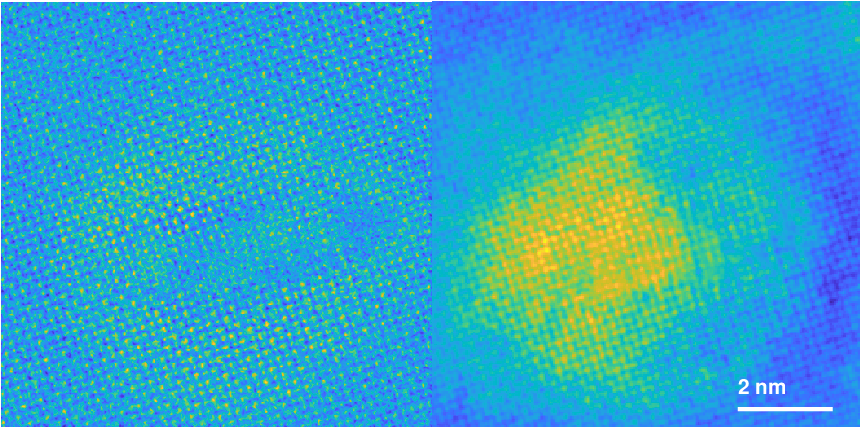}
  \caption{Comparison of iDPC (right) and Neural network reconstructions (left) of a FIB lamella, including a hole in the center. This illustrates reasonable tolerance of the proposed method towards thickness variations and bending.}
  \label{fig:sto_full}
\end{figure}

\begin{figure}[htb]
  \centering
  \includegraphics[width=\linewidth]{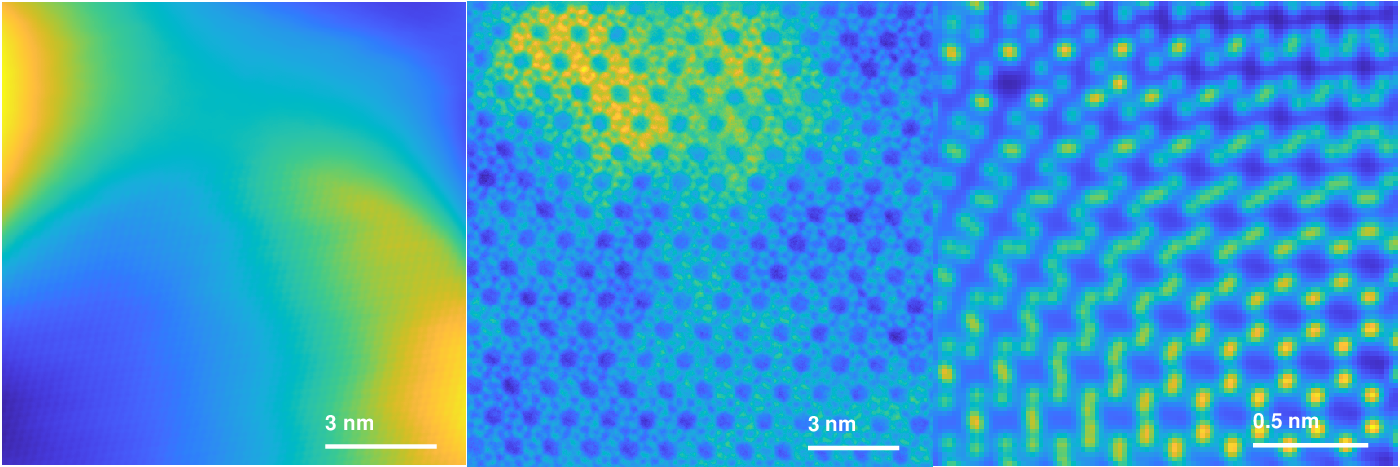}
  \caption{Additional iDPC reconstructions for comparison of samples presented in the paper, f.l.t.r: experimental Gold nanorod, experimental Zeolite, simulated twisted bilayer graphene with infinite dose.}
  \label{fig:idpc_all}
\end{figure}

\end{document}